\documentclass[preprint, prab, nofootinbib]{revtex4-2}

\usepackage[colorlinks=true, allcolors=blue]{hyperref}
\usepackage{graphicx}
\usepackage{dcolumn}
\usepackage{bm}
\usepackage{siunitx}
\usepackage{xcolor}
\usepackage{braket}
\usepackage{mathtools}
\usepackage{amsmath}
\usepackage{leftidx}
\usepackage{wasysym}
\usepackage{float}
\usepackage{cleveref}
\Crefname{equation}{Eq.}{Eqs.}

\usepackage{amsfonts} 
\DeclareMathSymbol{\shortminus}{\mathbin}{AMSa}{"39}

\newcommand{\dd}{\mathrm{d}}				

\graphicspath{{images/}}

\begin{document}


\title{Comment on ``Fast-slow mode coupling instability for coasting beams in the
presence of detuning impedance"}

\author{Alexey Burov}
\email{burov@fnal.gov}
\author{Valeri Lebedev}

\affiliation{Fermi National Accelerator Laboratory, Batavia, IL 60510, USA}%

\date{\today}

\begin{abstract}
In this comment we show untenability of key points of the recent article of N.~Biancacci, E.~Metral and M.~Migliorati [Phys. Rev. Accel. Beams 23, 124402 (2020)], hereafter the Article and the Authors. Specifically, the main Eqs.~(23), suggested to describe mode coupling, are shown to be unacceptable even as an approximation. The Article claims the solution of this pair of equations to be in ``excellent agreement" with the pyHEADTAIL simulations for CERN PS, which is purportedly demonstrated by Fig.~6. Were it really so, it would be a signal of a mistake in the code. However, the key part of the simulation results is not actually shown, and the demonstrated ``excellent agreement" has all the features of an illusion.             
          
\end{abstract}

\maketitle




The Authors stress that for a continuous beam, ``the effect of the detuning impedance is to add an additional tune shift to the bare machine working point". In other words, the detuning impedance is indistinguishable from the external quadrupole focusing. We agree with the Authors on that; our disagreement with them is about the relationship between beam optics and coupling of transverse collective modes. 

First, let us clarify the general coupling conditions and the terminology for the transverse modes of a coasting beam, since the Article~\cite{BiancacciPRAB2020} creates confusion in these respects. It is well-known that for a coasting beam, the transverse spectrum consists of two complex-conjugated series, which may be called {\it positive-based} and {\it negative-based}, $\Omega \approx (Q_\beta + n)\omega_0$ and $\Omega \approx (- Q_\beta + n) \omega_0$ respectfully; $n=0,\pm1,\pm2,...$ is an integer harmonic number. Driving impedance terms, which are relatively small, are dropped here for the moment, and the notations are same as in the Article. Each series, in turn, consists of four types of modes, distinguished by the signs and values of the angular phase velocity $\Omega/n$, see e.g. Ref.~\cite{lee2018accelerator}. For the positive-based modes, these types follow,

\begin{itemize}
\item{ Zero mode, $n=0$, $\Omega= Q_\beta \, \omega_0 \equiv \omega_\beta$;}
\item{ Fast mode, $n >0$, hence,  $\Omega/n > \omega_0$;}
\item{ Backward mode, $-Q_\beta <  n <0$, hence, $\Omega/n < 0$;}
\item{ Slow mode, $n<-Q_\beta$, hence, $0<\Omega/n < \omega_0$.}
\end{itemize}

Notation for the negative-based modes follows with complex conjugation, $n \rightarrow -n\,$, $Q_\beta \rightarrow -Q_\beta$,

\begin{itemize}
\item{ Zero mode, $n=0$, $\Omega=-Q_\beta \omega_0 \equiv -\omega_\beta$;}
\item{ Fast mode, $n<0$, hence,  $\Omega/n > \omega_0$;}
\item{ Backward mode, $0<n<Q_\beta$, hence, $\Omega/n < 0$;}
\item{ Slow mode, $n>Q_\beta$, hence, $0<\Omega/n < \omega_0$.}
\end{itemize}

\begin{figure*}[tbh!]
  \centering
  \includegraphics*[width=0.7\textwidth]{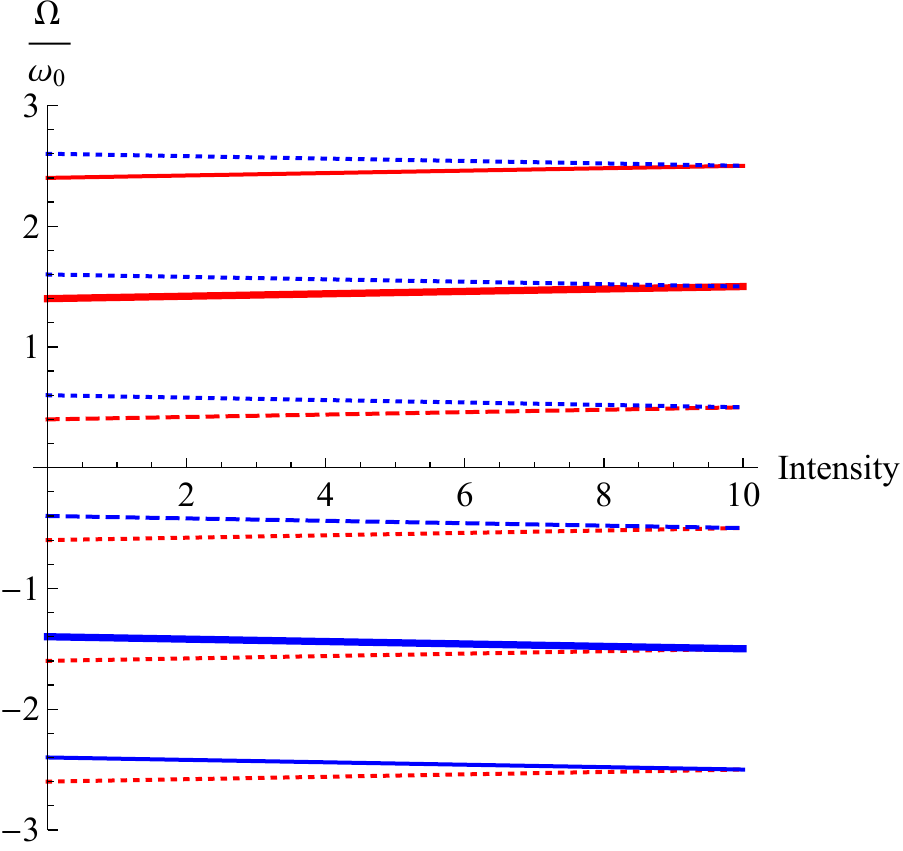}
  \caption{\label{PlotFastSlowModes} Sketch of a coasting beam spectrum, assuming smooth approximation and focusing detuning impedance. For graphical reasons, the lattice tune $Q_\beta=1.4$ is assumed, for this Figure only. Positive-based part is in red; negative-based is in blue. Zero modes are shown by thick solid lines, backward modes -- by dashed lines, slow ones -- by dotted lines, and fast modes -- by normal solid lines. All crossing lines necessarily have the same difference of the harmonic numbers, which is the nearest integer above the doubled tune, $n_1-n_2= \pm \mathrm{ceil}(2Q_\beta)=\pm 3$ in this example.}
\end{figure*}

Due to the driving impedance properties, only the slow modes can be unstable. An illustrative sketch of the spectrum is presented in Fig.~\ref{PlotFastSlowModes}, assuming smooth approximation and focusing detuning impedance, when the modes can cross but not couple. 

For the Article's PS example with lattice tunes $Q_\beta =\omega_\beta/\omega_0=6.4$, a slow mode with frequency $\omega_\beta -7 \omega_0 = -0.6 \omega_0$ has the nearest mode $-\omega_\beta +6 \omega_0 = -0.4 \omega_0$, the backward one. The Article, including the title, calls the latter mode ``fast," which is a terminological mistake: the value of the phase velocity of that allegedly ``fast" mode is actually smallest among all the modes.

In linear systems with time-independent coefficients, modes can couple only when their frequencies coincide. Clearly, positive-based modes can couple only with the negative-based ones, and vice versa; one of the modes must be stable (fast, zero, or backward), and another unstable (slow). For the PS example, the positive-based slow mode with $n=n_1=-7$ might couple with the negative-based backward mode with $n=n_2=+6$. The coupling can happen if the lattice tune difference between the two modes, $0.6-0.4=0.2$, is compensated by the detuning impedance, presumably able to shift the betatron tune up by $0.1$. Note that the difference between the harmonic numbers of the coupled modes, $n_2 - n_1 = 13 = \mathrm{ceil}(2Q_\beta)$, is just above the doubled betatron tune; the same is true for any pair of coupled modes. 

We beg pardon for this pedantic textbook explanation, but we feel obliged to make it since in the Article the terms are confused and the harmonic numbers are given without signs, making a false impression that the modes of the neighbor harmonics, 6 and 7, can sometimes be coupled.       

Let us now come back to Eqs.~(23). They are derived from Eq.~(22) by an ansatz that the collective oscillation $y(s,t)$ is a linear combination of two harmonics, $n_1$ and $n_2$. In that derivation, the dependence on $s$ was dropped without any explanation. For equations with constant coefficients, such as Eqs.~(23), this omission is a mathematical mistake, since the cross terms in Eqs. (23) depend on the coordinate $s$ in a different way than the direct terms, $y_1 \propto \exp(i n_1 s/R)$, while $y_2 \propto \exp(i n_2 s/R)$, and $n_1 \neq n_2$. To provide the mode coupling, the cross terms can only be built by harmonic $n_1 - n_2=\pm \mathrm{ceil}(2Q_\beta)$ of the driving impedance weighed with the beta-function, $Z^\mathrm{driv}(\Omega,s) \beta(s)$. In the smooth approximation, where $Z^\mathrm{driv}(\Omega,s) \beta(s) = \mathrm{const}$, these cross terms are equal to zero. Thus, instead of Eqs.~(23) of the Article, the mode coupling problem should be described by the following equations;
\begin{equation}
\begin{split}
& \Ddot{y}_1 + \omega_\beta^2 y_1 = -2\omega_\beta \Delta \Omega^\mathrm{tot}\, y_1 -2\omega_\beta \Delta \Omega^\mathrm{driv}_{n_1 - n_2} \,y_2 \,;	\\
& \Ddot{y}_2 + \omega_\beta^2 y_2 = -2\omega_\beta \Delta \Omega^\mathrm{driv}_{n_2 - n_1} y_1 - 2\omega_\beta \Delta \Omega^\mathrm{tot}\, y_2  \,.	\\
\end{split}
\label{TrueEq}
\end{equation}
Here $\Delta \Omega^\mathrm{tot} \propto i \int{ \dd s [Z^\mathrm{det}(0,s) + Z^\mathrm{driv}(\Omega,s)] \beta(s)}$ is the conventional uncoupled coherent tune shift at the sought-for frequency $\Omega \approx \pm \omega_\beta +n \omega_0$, and the cross-coefficients can be expressed as 
\begin{equation}
\Delta \Omega^\mathrm{driv}_{n_1 - n_2} = \Delta \Omega^\mathrm{driv}_0 
\frac{\int{ \dd s\, Z^\mathrm{driv}(\Omega,s) \beta(s) \exp(i (n_1 - n_2)s/R)}}{\int{ \dd\, s Z^\mathrm{driv}(\Omega,s) \beta(s)}}\,,
\label{CrossCoeff}
\end{equation}
with $\Delta \Omega^\mathrm{driv}_0$ as the contribution of the driving impedance into the conventional coherent tune shift, $\Delta \Omega^\mathrm{driv}_0 \propto i \int{ \dd s\, Z^\mathrm{driv}(\Omega,s) \beta(s)}$. Let us stress again, that within the smooth approximation, apparently presumed by the Article, but contrary to its Eqs.~(23), the cross coefficients can only be zeros, since the integral in the numerator of Eq. (2) is equal to zero for $n_1 \neq n_2$. It is also worth noting, that the mode coupling described by Eq.~(\ref{TrueEq}) can hardly be of practical importance: the coupling may show itself only near the half-integer resonance, wherefrom the tunes should be kept out anyway. 

The Authors claim that the proposed instability mechanism is conceptually analogous to the transverse mode coupling instability (TMCI) for bunched beams. Here they overlook an important difference between the azimuthal harmonics of a coasting beam and synchrotron harmonics of a bunch. The former are exact eigenfunctions for the coasting beam for arbitrary impedance, provided that the smooth approximation is justified, which is apparently assumed in the Article. This fact is guaranteed by the translation invariance of the dynamic equations. Contrary to that, the synchrotron harmonics of a bunch can, at best, be only approximations for the eigenfunctions, which accuracy deteriorates when the coherent tune shift becomes comparable with the synchrotron tune. For a bunch, the dynamic equations are not invariant under the synchrotron phase shifts, and thus, strictly speaking, the synchrotron harmonics can constitute the eigenfunctions only at zero wake field. That is why coupling of the azimuthal harmonics is forbidden for homogeneous coasting beams in the linear approximation, while coupling of the synchrotron harmonics of a bunch is possible. Note also that the eigenvectors of equidistant multiple bunches are the same harmonic exponents, for any impedance.  
 
Now we are coming to the last point of this comment, to the claimed excellent agreement between the theory of Eqs.~(23) and pyHEADTAIL simulations for the PS presented in Fig.~6. Were this agreement really there, it could only mean that a mistake occurs also in the code. However, we think that Fig.~6 demonstrates something different. Rather, it demonstrates an illusion of agreement. To show this, let us first have a look on the upper plot of this Figure, which represents the growth time $\tau$[turns]. Here, we clearly see an agreement for all the points with $\tau \gg 1$, where the cross terms of Eqs.~(23) do not make any visible difference. For these points we indeed see agreement between the textbooks and the pyHEADTAIL simulations, while nothing at all can be said about the innovative aspect of the Article. Apart from these conventional points, we also see in this plot a sequence of points and the line at the growth time $\tau \approx 0$. These results of the code and Eqs.~(23) might really demonstrate the agreement or disagreement between the theory and the simulations, but for that we have to see these data. Instead, we see something indistinguishable from zero, or from infinity, if to speak in terms of the growth rates. Thus, the data of the upper plot of Fig.~6 consist of two parts: one is irrelevant to the suggested model of the mode coupling, and the other is obscured from judgement about the model validity by means of the way the data are presented. 

As to the bottom plot of Fig.~6, we see there that the mode tunes are locked in the half-integer resonance. For the simulations, something like that has to be expected just on the ground of the sufficiently strong detuning quadrupole for a lattice with nonzero harmonic $13$. For a perfectly smooth lattice, however, the half-integer tune $6.5$ would be as good as any other tune; thus, the result of the simulations must be sensitive to the lattice smoothness. Since Eqs.~(23) are fully insensitive to the phase advance per cell or other smoothness parameters, the agreement between the pyHEADTAIL simulations and theory in the bottom plot of Fig.~6 can be only accidental.  

We'd like also to note that, although the half integer resonance is not presented in the smooth approximation, it plays a significant role in any real machine. Approaching this resonance results in a big variation of beta-functions and, consequently, fast increase of effective impedance $ \left< Z^\mathrm{driv}(\Omega,s) \beta(s) \right >$ and its coupling-related harmonic, mentioned above. 

We hope that our disagreement with key issues of the Article is clearly expressed, and we would appreciate a response of the Authors.


This manuscript has been authored by Fermi Research Alliance, LLC under Contract No. DE-AC02-07CH11359 with the U.S. Department of Energy, Office of Science, Office of High Energy Physics. 

\bibliography{bibliography} 

\end{document}